\documentclass[9pt,twocolumn,twoside]{osajnl}

\journal{ol} 

\setboolean{shortarticle}{true}
\newcommand{\p}{\partial}
\newcommand{\ep}{\varepsilon}

\newcommand{\om}{\omega}
\newcommand{\nn}{\nonumber}
\newcommand{\ta}{\theta}

\newcommand{\be}{\begin{equation}}                                              \newcommand{\ee}{\end{equation}}
\newcommand{\ba}{\begin{eqnarray}}
\newcommand{\ea}{\end{eqnarray}}
\title{Frequency combs in a microring optical parametric oscillator}
\author[1]{A.~Villois}
\author[2]{N.~Kondratiev}
\author[3,4]{I.~Breunig}
\author[1]{D.N.~Puzyrev}
\author[1,2,*]{D.V.~Skryabin}
\affil[1]{Department of Physics, University of Bath, Bath BA2 7AY, UK}
\affil[2]{Russian Quantum Centre, Skolkovo 143025, Russia}
\affil[3]{Department of Microsystems Engineering - IMTEK, University of Freiburg, 79110 Freiburg, Germany}
\affil[4]{Fraunhofer Institute for Physical Measurement Techniques, 79110 Freiburg, Germany}
\affil[*]{Corresponding author: d.v.skryabin@bath.ac.uk}
\dates{Compiled \today}
\ociscodes{(190.4975)   Nonlinear optics,  Parametric processes; (140.3945) Microcavities; (190.5530)   Pulse propagation and temporal solitons; (190.1450)   Bistability}
\begin{abstract}
We report  the soliton frequency comb generation in microring optical parametric oscillators operating in the down-conversion regime and with the simultaneous presence of the  $\chi^{(2)}$ and Kerr nonlinearities. The combs are studied considering a typical geometry of a bulk LiNbO$_3$ toroidal resonator with the normal group velocity dispersion spanning an interval between the pump and the down-converted signal.  We have identified critical power signaling a transition between the relatively low pump power predominantly $\chi^{(2)}$ combs and the high pump power ones shaped by the competition between the $\chi^{(2)}$ and Kerr nonlinearities.
\end{abstract}

\setboolean{displaycopyright}{true}

\begin{document}
	
	\maketitle
	
		\begin{figure}[t]
		\centering
		\includegraphics[width=0.49\textwidth]{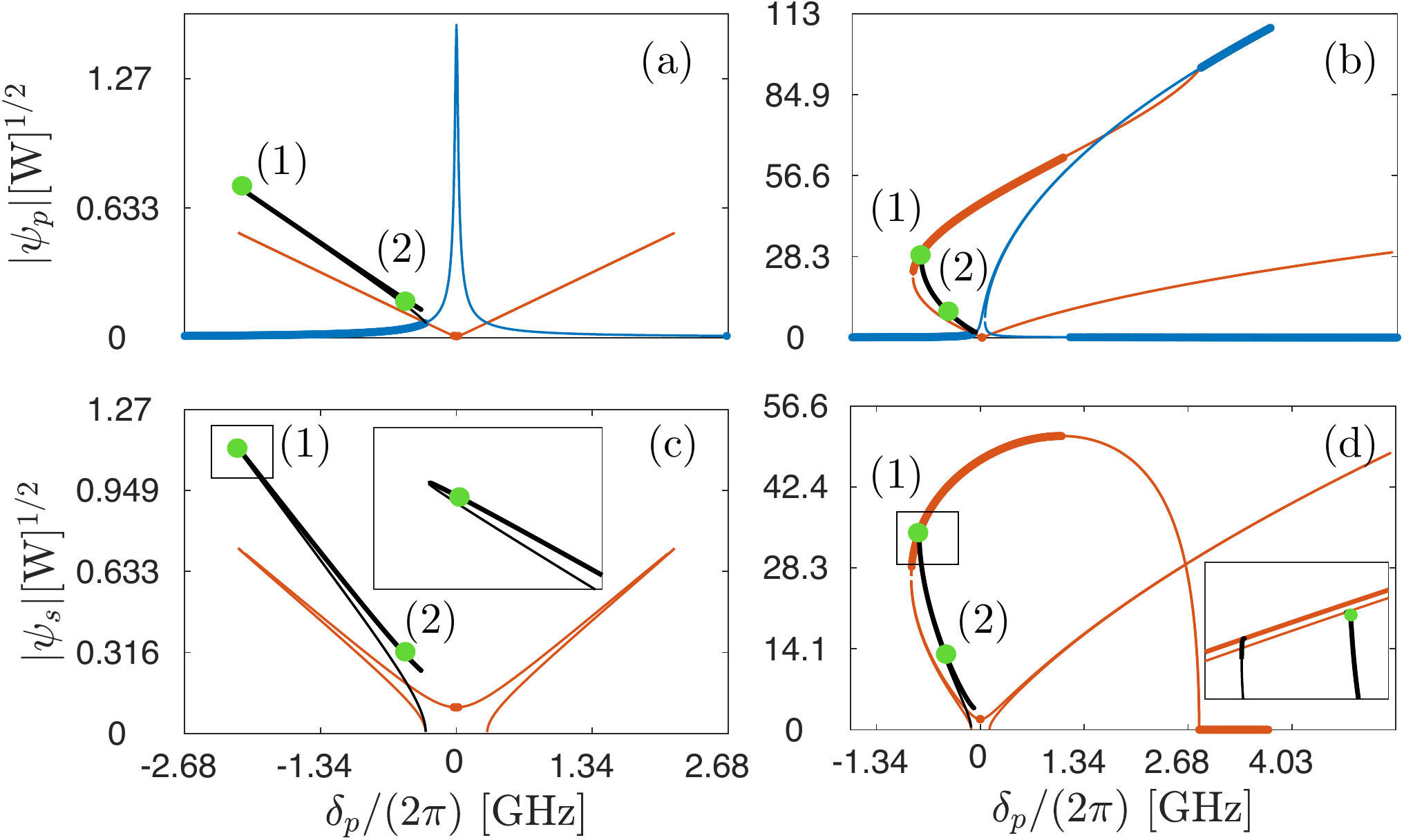}
		\caption{ 
			Pump (a,b) and signal (c,d) amplitudes of the cw states vs detuning.
			Parametric effect is dominant in (a,c) ($Q=10^7$, $P=10$mW$\ll P_{cr}$, $\chi_2/\chi_3=5\cdot 10^{9}$V/m). Kerr and parametric effects compete in (b,d) ($Q=10^8$, $P=2.5$W$\sim P_{cr}$, $\chi_2/\chi_3=5\cdot 10^7$V/m). Blue/red lines show the pump-only/parametric cw states.  
			Black lines denote maxima of the intensities of the bright parametric solitons. Zooms for the rectangular areas in (c) and (d) are shown in the insets. Thinner lines mark unstable cw states and solitons.
		}\label{fig1}
	\end{figure}
Frequency combs are a key concept and a tool in the fields of short pulse generation,  optical metrology, and spectroscopy \cite{comb1}. The relatively recent development of the high-quality factor whispering gallery mode (WGM) microring resonators opened up doors for the chip and micro scale frequency comb (microcomb) devices \cite{comb2}. Association of the microcombs with the soliton solutions of the Lugiato-Lefever (LL) equation \cite{Herr2013} has led to the emergence of a burgeoning research area of the microring solitons supported by the interplay of the positive Kerr nonlinearity and anomalous group velocity dispersion (GVD) \cite{Kippenberg2018}. Reducing power thresholds for the soliton comb and extending their range of existence to the normal dispersion are often desirable for practical applications and therefore quadratic nonlinear response, $\chi^{(2)}$,  of the noncentrosymmetric materials  is a viable exploration direction in the microcomb area that has been recently studied by a number of authors \cite{Hansson2018,Villois2019,hans2,Herr2,ibm,vahala,china,ikuta}. Importantly, the comb solitons embracing 2nd harmonic are naturally octave wide and therefore suitable for self-referencing \cite{comb1,comb2}.  Among the possible candidate materials to study $\chi^{(2)}$ soliton combs, the lithium-niobate (LN) results are particularly promising given the large $\chi^{(2)}$ coefficient and its transparency over a wide spectral window. 
	Moreover, recent developments in LN nanophotonics has shown its potential for the broadband frequency conversion, dispersion engineering, quasi-phase matching, and straightforward  on-chip integration, see, e.g., \cite{Wang2018,ing}. Note, that the area of quadratic comb solitons connects to the prior research on the  quadratic solitons \cite{skr} and parametric conversion in microresonators \cite{ingo}.
	
	While the soliton combs associated with the second harmonic generation process in microring resonators have received some preliminary attention in the recent publications \cite{Hansson2018,Villois2019,china,ikuta},  we demonstrate below the parametric down-conversion is equally promising for the comb generation.   We consider a typical geometry of the LiNbO$_3$ toroidal microresonator having the normal GVD range between $800$nm and $1600$nm and study formation of the frequency combs when  the system operates in the parametric down-conversion regime. We account for both $\chi^{(2)}$ and $\chi^{(3)}$ nonlinearities and demonstrate how their competition may alter the comb generation.

 Recent experiments with a microring pumped with the external comb signal with the matched repetition rate  have reported the line by line conversion into multiple higher and lower frequency bands  \cite{Herr2}.  Also recent publications on the meter scale bow-tie cavities loaded with a $\chi^{(2)}$ crystal and operating in the down-conversion regime have reported a low-coherence non-soliton combs with very few-side bands due to GVD induced instabilities \cite{mosc} and the so-called simulton-solitons supported solely by the group velocity mismatch, where GVD plays no role \cite{martin}. While the coherent multi-side-band microring parametric combs reported below rely on the normal GVD and thus are the first of its kind.
		\begin{figure}[t]
		\centering
		\includegraphics[width=0.49\textwidth]{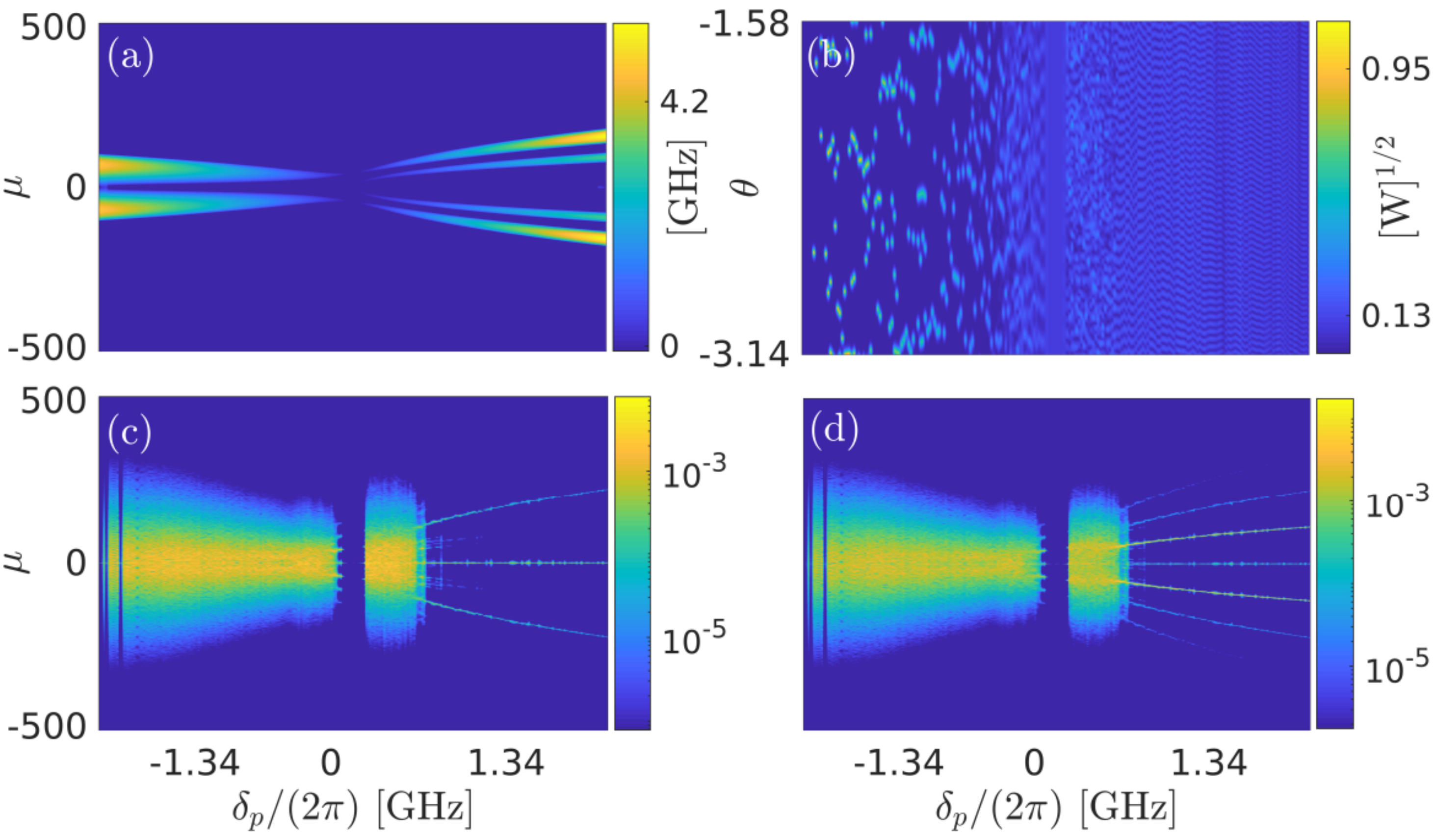}
		\caption{
			(a) MI growth rate $Re\Lambda$ vs $\delta_p=2\delta_s$ and the modal index $\mu$ for the high amplitude parametric cw state. 
			(b) Spatial distribution of the signal amplitude after a 
			half-million of round trips vs $\delta_p$. $\theta$ is shown between $-\pi/2$ and $-\pi$ for better visibility. (c)/(d) Log-scale modal spectrum of the pump/signal after a half-million of round trips vs $\delta_p$. 
			All parameters are as in Figs. 1(a) and 1(c), i.e., $P\ll P_{cr}$.
			\label{fig2}}
	\end{figure}	

Since the derivation of a model for the comb generation in a microring resonator with the $\chi^{(2)}$ nonlinearity is not readily available in the literature, we outline a version of it below and map it onto the practical parameters of an LN microring. We start from the averaged along the transverse resonator cross section scalar wave equation $\p_z^2E-c^{-2}\p_t^2\int_{-\infty}^{\infty}g(t-t')E(t')dt'=\mu_0\p^2_t\langle{\cal P},f\rangle/\langle f,f\rangle$.
	Here $z=R\ta$ is the coordinate along the ring, $R$ is the ring radius,  $\ta$ is the azimuthal angle, $fE$ is the electric field  and $c^{-2}=\mu_0\ep_0$. 
	$f$ is the dimensionless transverse modal profile with the   modal area $S\simeq 30\mu$m$^2$.
	$\langle \cdot,\cdot\rangle$ means integration over the transverse dimensions of the modal profile.
	$g$ is the resonator response function accounting for a combination of the material and geometrical dispersions, such that $\int_{-\infty}^{\infty}g(t)e^{-i\om t}dt=n_e^2(\om)$, where $n_e(\om)$ is the  effective refractive index.  ${\cal P}=\ep_0\chi_2 f^2E^2+\ep_0\chi_3 f^3E^3$ is the nonlinear polarization combining quadratic, $\chi_2\simeq 5\cdot 10^{-12}$m/V, and cubic, $\chi_3\simeq 10^{-21}$m$^2$/V$^2$, nonlinear susceptibilities of LN.  $f$ and $\chi_{2,3}$ are assumed non-dispersive for the sake of brevity and transparency of notations.  For ${\cal P}=0$, $E\sim e^{im\ta-i\om_mt}$, the linear resonator spectrum is recovered from $c^{2}m^2=R^{2}\om_m^2n_e^2(\om_m)$ . Taking $E=b\sum_m A_me^{im\ta-i\om_mt}$, ${\cal P}=\sum_m {\cal P}_me^{im\ta-i\om_mt}$, assuming that the Fourier amplitudes vary slowly in time and dispersion and nonlinearity are relatively small, we find $ib\p_tA_m=-\frac{\om_m}{2\ep_0n^2}\langle {\cal P}_m,f\rangle/\langle f,f\rangle$,
	where $n\simeq 2.2$. Here $b^2=2/[\ep_0cnS]\simeq  10^{13}$V$^2$m$^{-2}$W$^{-1}$ and, thereby  $|A_m|^2$ is measured in Watts.  The  above procedure renders explicit expressions for the nonlinear coefficients in the LL model, see $\gamma_{2,3}$ below. 
	
The derivation should be applied for the both extraordinary (polarised along the resonator axis) and ordinary (polarised in the resonator plane) waves, that  appropriately couple through the nonlinear polarization tensor. 
We now assume that the extraordinary pump (p) and ordinary signal (s) waves have their own modal profiles $f_{p,s}$ and their electric fields are given by  $b_{p,s}f_{p,s}\psi_{p,s}(t,\ta)e^{im_{p,s}\ta-i\om_{p,s} t}+c.c.$, where $\psi_{p,s}$ are the intracavity pump and signal amplitudes spectrally capturing a sufficiently large number of the resonator lines and $b_{p,s}^2=2/[\ep_0cnS_{p,s}]$ are the respective power scaling coefficients. $\omega_p$ is the cavity resonance nearest to the pump frequency $\Omega$ and $\omega_s$ is the one nearest to $\Omega/2$.  Then transforming from the modal expansion into the real space as per, e.g, \cite{kkk}, including the pump field with frequency $\Omega$ nearest to $\om_p$ and losses we obtain the  LL-like system 
\ba
	&& \nn i\p_t\psi_s=\left(\delta_s-iD_{1s}\p_\ta-\tfrac{1}{2}D_{2s}\p^2_\ta\right)\psi_s-i\kappa_s\psi_s\\  && -\gamma_{2s} \psi_s^*\psi_p e^{i2\ep t+i(m_p-2m_s)\ta}
	-\gamma_{3s}(|\psi_s|^2+2a_s|\psi_p|^2)\psi_s,\label{e1}\\
	&&\nn i\p_t\psi_p=\left(\delta_p-iD_{1p}\p_\ta-\tfrac{1}{2}D_{2p}\p^2_\ta\right)\psi_p-i\kappa_p\psi_p+h\\ && -\gamma_{2p}\psi_s^2e^{-i2\ep t-i(m_p-2m_s)\ta}-\gamma_{3p}(|\psi_p|^2+2a_p|\psi_s|^2)\psi_p.
	\label{e2}\ea
Here $\delta_s=\om_s-\Omega/2$ and $\delta_p=\om_p-\Omega$ are the detuning parameters. 
In what follows we assume that $\chi_2$ varies periodically in $\ta$, so that a mismatch of the modal indices, $m_p-2m_s$, is compensated in the leading order \cite{ingo,ing}, while the higher order corrections are disregarded.
$\ep=\om_s-\om_p/2$ is the frequency mismatch parameter. 
Note, that $\delta_s=\delta_p/2$ for $\ep=0$, which is assumed in our numerics. $D_{1p,1s}$  are the local free spectral ranges (FSRs)  and $D_{2p,2s}$ set the GVD associated FSR  dispersion.   $\gamma'$s are the quadratic and cubic nonlinear coefficients. 
In what follows we assume for brevity $\gamma_{2s,2p}\simeq\gamma_2=
\frac{\om_{p}\chi_2 N_{2p}b_p}{2n^2}$, $\gamma_{3s,3p}\simeq\gamma_3=\frac{\om_{p}\chi_3 N_{3p}b_p^2}{2n^2}$, where $N_{jp}=\int f_{p}^{j+1}dS[\int f_{p}^2dS|]^{-1}$. For the parameters specified above $\gamma_2\simeq 9\cdot 10^9$W$^{-1/2}$s$^{-1}$ and $\gamma_3\simeq 10^6$W$^{-1}$s$^{-1}$. 
$h=\sqrt{\eta\kappa_pD_{1p}P/(2\pi)}$ is the pump parameter, where $P$ is the  pump power  \cite{kkk}. 
The quality factor of LN microresonators strongly depends on the chosen platform. The chip integrated samples are known to reach $Q\simeq 10^7$ \cite{lon}. However, the bulk  LN resonators can have a better $Q$. E.g., the ones used in \cite{opex} have had extinction coefficients $r\simeq 5\cdot 10^{-4}$cm$^{-1}$, which corresponds to the intrinsic quality factor
$2 \pi n / (r\lambda)\simeq 2\cdot 10^8$ in the required wavelength range, $\lambda$.  Total $Q$ is half of the above for the critical coupling ($\eta=1/2$), and thus for the moderate $Q=10^7$, we have $2\kappa_{p,s}\simeq\om_{p,s}/Q\sim 10^8$s$^{-1}$. 
	\begin{figure}[t]
		\centering
		\includegraphics[width=0.49\textwidth]{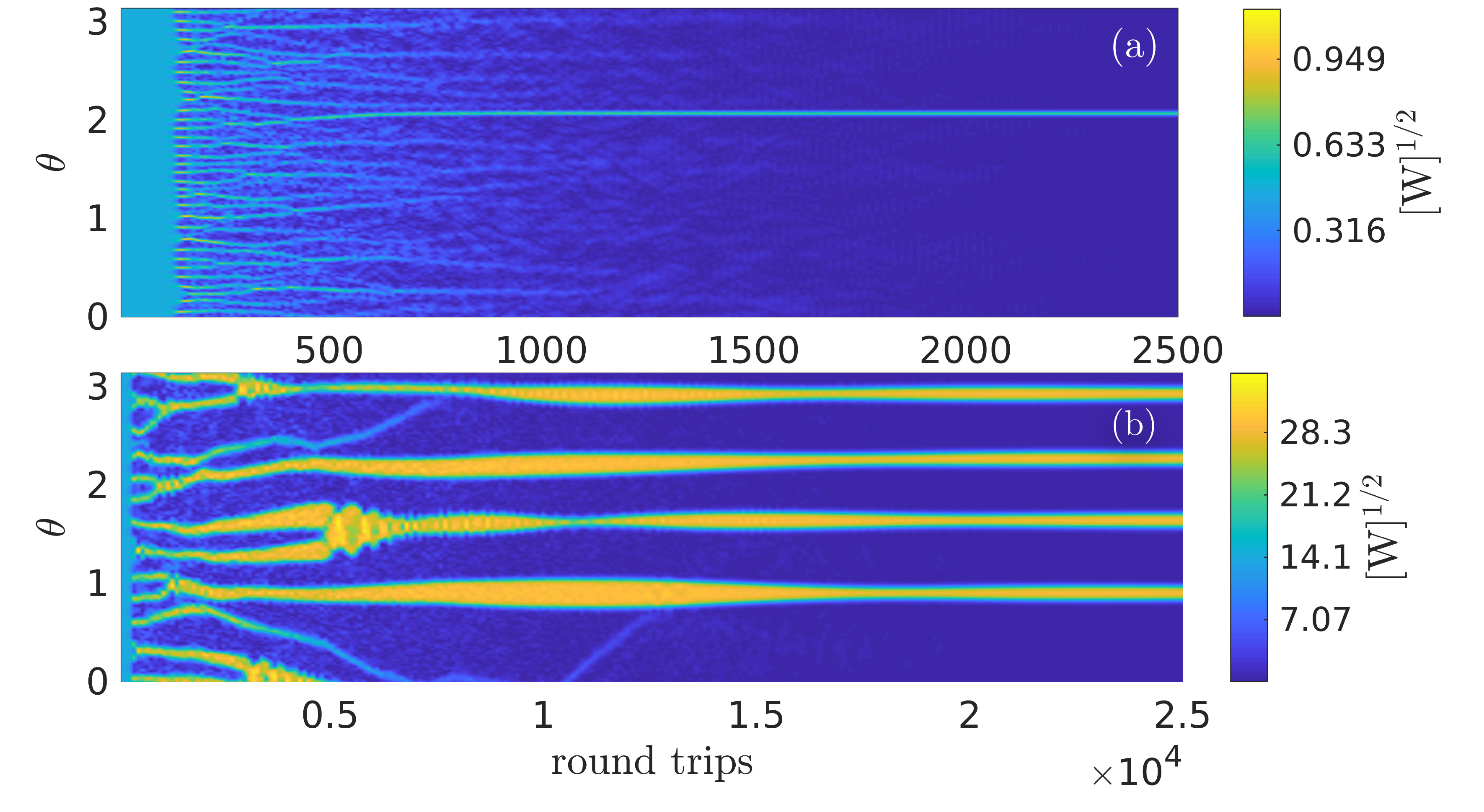}
		\caption{Time-space evolution of the signal amplitudes when the cavity is initialised with an unstable cw state for $P\ll P_{cr}$ (a) and $P\sim P_{cr}$ (b).  Parameters in (a)/(b) are as in Fig. 1(a)/1(b) with $\delta_p/(2\pi)=-1.97$/$-0.79$GHz. $\theta$ is shown only between $0$ and $\pi$ for better visibility.}\label{fig3}
	\end{figure}
	
To determine practical FSRs and GVDs we consider a spheroidal LN microresonator (major radius $R\sim500~\mu m$, minor radius $r\sim250~\mu m$) and use analytical expressions for the resonance frequencies \cite{dem}. By varying $r$ and $R$ we found a pair of resonances $\omega_p$ (extraordinarily in the proximity of $0.8~\mu m$) and $\omega_s$ (ordinary in the proximity of $1.6~\mu m$) such that the FSR parameters $D_{1p}$ and $D_{1s}$ are approximately matched, and, simultaneously,  the frequency mismatch parameter $\ep$
is reduced to nearly zero. Close matching of FSRs is also known as an important condition to achieve soliton combs via second harmonic generation \cite{Villois2019}. Choosing $\omega_p/(2\pi)=381.749381$~THz (azimuthal mode number $m_p=8668$) gives $\omega_s/(2\pi)=190.874707$~THz (azimuthal mode number $m_s=4401$). Then $\ep\simeq 0$, $D_{1s}\simeq D_{1p}=2\pi\cdot 42.17$~GHz, and  $D_{2s}=-2\pi\cdot 166.94$~kHz, $D_{2p}=-2\pi\cdot 548.30$~kHz. Non-zero values of $\ep$ and $1-D_{1s}/D_{1p}$ do not qualitatively alter the results described below.

The nonlinear resonances of Eqs. (\ref{e1}), (\ref{e2}) are characterised by their continuous wave (cw) solutions ($\p_t=\p_\ta=0$). In the down-conversion setting, there exists a cw  state with the nontrivial pump, $\psi_p\ne 0$, and, the zero signal, $\psi_s= 0$. This pump-only state exists both above and below the  threshold where the parametric state, $\psi_{p,s}\ne 0$,  splits away from $\psi_s= 0$. The threshold condition,  $h^2\gamma_2^2=(\delta_p^2+\kappa_p^2)(\delta_s^2+\kappa_s^2)$, can be found explicitly for $\gamma_3=0$. The blue and red lines in Fig. 1 mark the pump-only and the parametric cw states, respectively. The resonance line of the pump-only state tilts towards  positive values of $\delta_p$  through the Kerr effect induced bistability. By approximately solving the Kerr bistability equation, $h^2=(\kappa_p^2+(\delta_p-\gamma_3|\psi_p|^2)^2)|\psi_p|^2$, in the limit of large detunings, we find that the extent of the tilt  in $\delta_p$ is approximated as  $\delta_{p}^{Kerr}=\gamma_3h^2/\kappa_p^2(1+{\cal O}(h\gamma_3^{1/2}/|\delta_p|^{3/2}))$.  

In the presence of the parametric effect, the linear cavity resonance splits into two resonances oppositely tilted towards both negative and positive detunings, see Fig. 1, which is similar to what was previously described for the second harmonic generation in microrings \cite{Hansson2018,Villois2019}. Thus the parametric effect acts similarly to the positive/negative Kerr nonlinearity  for $\delta_p>0$/$\delta_p<0$. For $\gamma_3=0$, the extent of the parametric tilt can be found from the exact condition $h^2\gamma_2^2=(\kappa_s\delta_p+\kappa_p\delta_s)^2$, so that for $\ep=0$ and $\kappa_p=\kappa_s$, we get $\delta_p^{prm}=\pm 2\gamma_2h/[3\kappa_p]$.
While the width of the parametric resonance in $\psi_s$ field is proportional to $\kappa_s$,  the resonance line in the $\psi_p$ component has a much smaller, even vanishingly small (for $\gamma_3=0$), width, cf. Figs. 1(a) and 1(c). 
	\begin{figure}[t]
		\centering
		\includegraphics[width=0.49\textwidth]{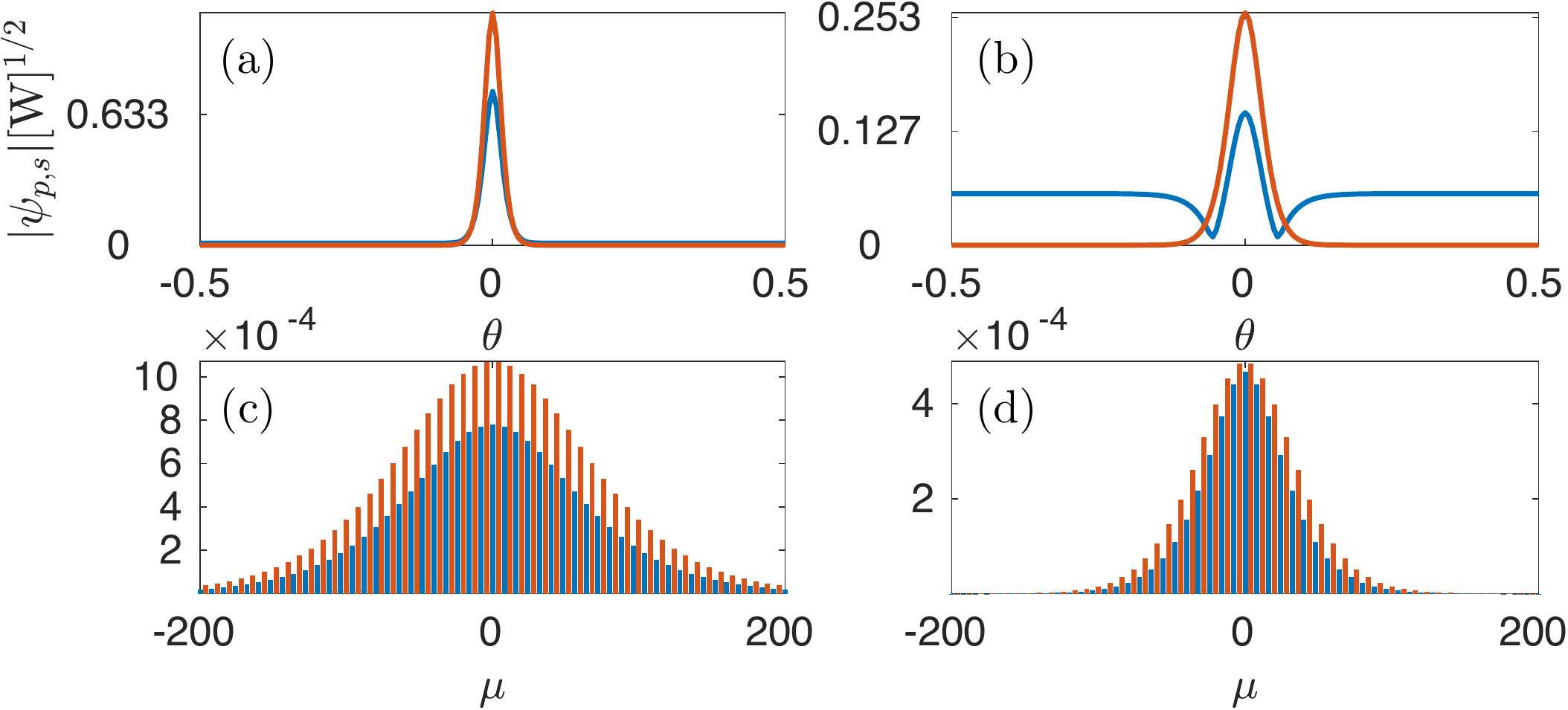}
		\caption{Parametric soliton profiles (a,b) and the corresponding comb spectra (c,d) for $P\ll P_{cr}$. (a,c)/(b,d) correspond to the points 1/2 in Fig. 1(c). Blue and red colors correspond to the pump and signal, respectively.
		} \label{fig4}
	\end{figure}
	\begin{figure}[t]
		\centering
		\includegraphics[width=0.49\textwidth]{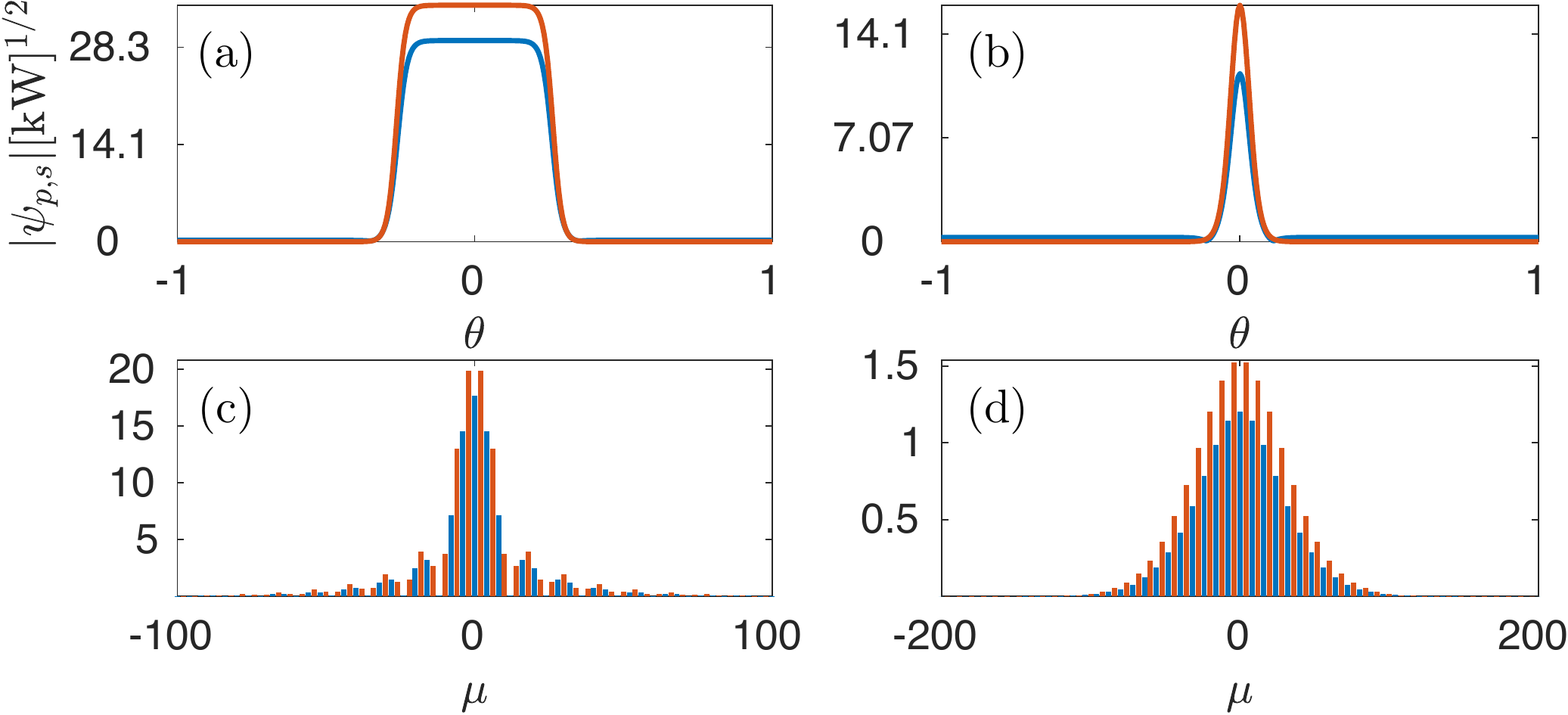}
		\caption{Same as Fig.4, but for $P\sim P_{cr}$. (a,c)/(b,d) correspond to the points 1/2 in Fig. 1(d).
		} \label{fig5}
	\end{figure}
	
Thus the nonlinear shift of the resonance frequency is proportional to $h^2$ for the Kerr effect and to $h$ for the parametric one. Hence the latter dominates for the relatively low pump powers, when $|\delta_p^{prm}|\gg\delta_p^{Kerr}$, while the two processes start to compete when $|\delta_p^{prm}|\sim\delta_p^{Kerr}$, i.e., at $h\sim h_{cr}=2\gamma_2\kappa_p/[3\gamma_3]$, which corresponds to $P\sim P_{cr}$, where
\be P_{cr}= \frac{4\pi\om_p}{9\eta QD_{1p}}\frac{\gamma_2^2}{\gamma_3^2}=\frac{4\pi}{9\eta {\cal F}}\cdot\frac{\chi_2^2}{b^2\chi_3^2}\cdot\frac{N_{2p}^2}{N_{3p}^2},\ee where ${\cal F}=QD_{1p}/\om_p$ is the cavity finesse. Using parameters as in the above text and $Q=10^7$ we find $P_{cr}\simeq 70$kW. By taking a higher, but still realistic, $Q=10^8$ and a material with  $\chi_2/\chi_3=5\cdot 10^7$V/m, instead of previous $5\cdot 10^{9}$V/m, we find a very practical $P_{cr}\simeq 1$W. The oppositely tilted parametric resonances are expected to lose their symmetry for $P$ approaching $P_{cr}$, see Figs. 1(b) and 1(d). For the positive detunings the parametric and Kerr effects  impact the resonance in the same way by extending its range towards larger positive detunings. However, for the negative detunings   nonlinearities, competing nonlinearities reshape the negatively tilted resonance by reversing the tilt direction. The signal part of the folded back solution  hits $\psi_s=0$ at some positive $\delta_p$, while the corresponding $\psi_p$ naturally merges with the pump-only cw state.  	
For the  few Watts pump power, one will certainly face thermally induced frequency shifts due to absorbed intracavity light, which  should be addressed in future studies.

	To study stability of the cw solutions with respect to the side-band excitation (modulational instability, MI), we make an ansatz $\psi_{p,s}(\theta,t)=\psi_{p,s}+(u_{p,s}(t,\theta)+iv_{p,s}(t,\theta))$, where $u_{p,s},v_{p,s}$ are small perturbations. Solving the linearized problem for the latter numerically we find the side-band growth rates $Re\Lambda$ as a function of their modal indexes $\mu$:  $u_{p,s}$, $v_{p,s}\sim e^{i\mu\theta+\Lambda t}$.   Results of this  analysis are shown in Fig. 1 using thick/thin  lines for  the stable/unstable cw states. Fig. 2(a) shows the MI growth rate of the parametric $P\ll P_{cr}$ cw state with the highest amplitude as a function of $\delta_p=2\delta_s$ and of the modal index $\mu$. The MI structure remains qualitatively similar for $P\sim P_{cr}$.

	To check the outcomes of the MI development we have performed an extensive series of dynamical  simulations. Fig 2(b) shows the spatial distributions of the signal field intensity, $|\psi_s|^2$,  after a half-million cavity round trips for a broad interval of the detunings and for $P\ll P_{cr}$.  Small amplitude and spectrally narrow waves are generated in the system for the large positive $\delta_p$, see Figs. 2(b)- 2(d).  More modes and broad, albeit non-solitonic, frequency combs are excited and the system exhibits complex spatio-temporal dynamics characterized by  the high amplitude oscillations for $\delta_p$ approaching zero. When $\delta_p\sim 0$, the cw solutions becomes stable in a narrow interval of $\delta_p$. For the negative values of $\delta_p$ the system enters the soliton regime, see well localised spots in Fig. 2(b). Note, that the initial noise realisations applied vary as $\delta_p$ is scanned. Fig. 3  shows the spatio-temporal dynamics of the signal field component leading to the formation of multiple solitons from the cw input. For $P\sim P_{cr}$, see Fig. 3 (b), the solitary filaments exhibit a strong breathing dynamics associated with the repulsion and attraction of the steep wave fronts connecting the low and high amplitude cw states. With more round-trips, the breathers are gradually converging to the multi-soliton state.
		\begin{figure}[t]
		\centering
		\includegraphics[width=0.49\textwidth]{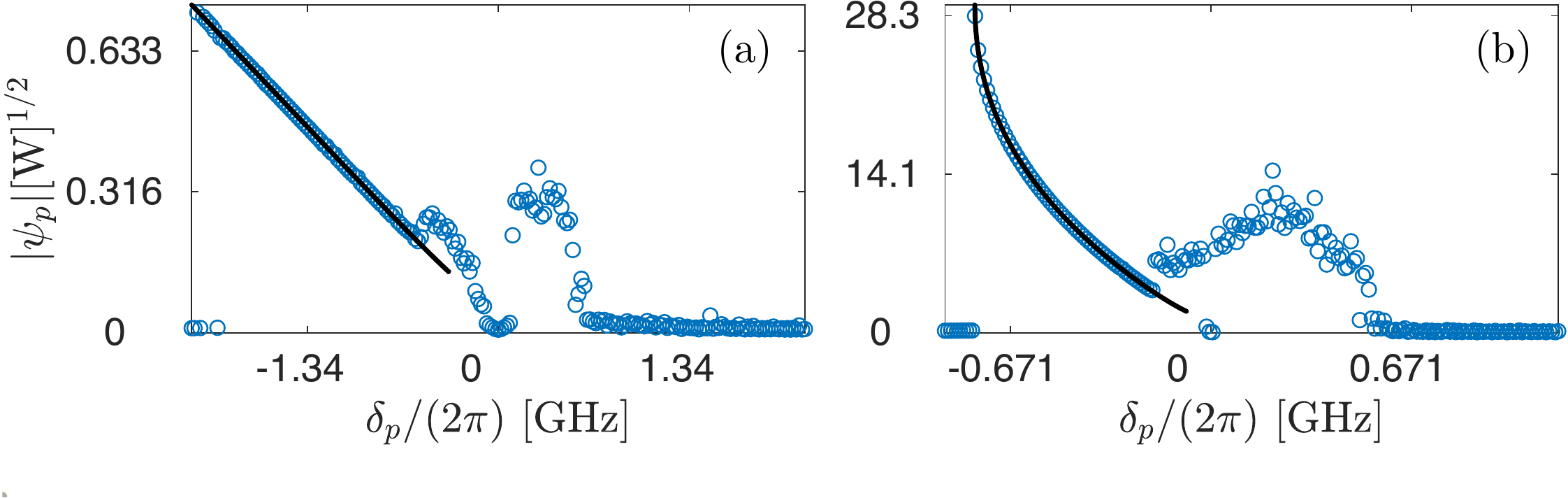}
		\caption{Comparison between the pulse amplitudes of the pump field obtained in the dynamical modelling (blue circles) with the stationary soliton amplitudes (black lines) for the $P\ll P_{cr}$ (a) and $P\sim P_{cr}$ (b) cases. Parameters are as in Figs. 1 and 2. }\label{fig6}
	\end{figure}
	
To find stationary soliton solutions we have numerically integrated Eqs. (\ref{e1}) and (\ref{e2})  with $\p_t=0$ using a Newton-Raphson method in combination with a bi-conjugate gradient method. Fourier space  with 2048 modes was used so that the periodic boundary conditions were accounted for automatically. Maxima of the soliton amplitudes are plotted in Fig. 1 using solid black lines.  Soliton families were found to have the resonance shapes, similar to the cw ones.  The linear stability of the solitons was also analyzed numerically. The higher amplitude soliton branch is typically stable and the low amplitude one is typically unstable. Both nonlinearities provide positive shift of the cavity resonance for  $\delta_p=2\delta_s>0$, which does not favour bright solitons for the normal dispersions, $D_{2p,2s}<0$. If, however, $\delta_p=2\delta_s<0$ then the negative shift of the resonance by the quadratic nonlinearity prevails for  $P\ll P_{cr}$ and creates conditions for stable bright solitons. It is instructive to present an approximate analytical soliton solution, that can be found assuming $\gamma_3=D_{2p}=0$, $D_{1p}=D_{1s}$, $\kappa_p/\delta_p\ll 1$. Then for $\psi_p\delta_p\simeq \gamma_2\psi_s^2-h$, $x=\ta-D_{1p}t$,
	we find that $\psi_s=\psi e^{i\alpha}$, where $\psi$ is a solution of $\frac{1}{2}D_{2s}\p_x^2\psi=q_{\pm}\psi-\gamma_2^2\psi^3/\delta_p$ and $\sin 2\alpha=-\kappa_s\delta_p/(\gamma_2h)$  \cite{barash}. Thereby 
	\be
	\psi_s=\frac{e^{i\alpha}}{\gamma_2}\sqrt{2\delta_p q_{\pm}}\cdot sech~x\sqrt{\frac{2q_{\pm}}{D_{2s}}}, ~q_{\pm}=\delta_s\pm\sqrt{\frac{\gamma_2^2h^2}{\delta_p^2}-\kappa_s^2}.\label{e4}\ee
	Eq. (\ref{e4})  provides a good qualitative understanding of the parametric soliton combs. It explicitly predicts   that the bright comb solitons  exist for normal GVD, $D_{2s}<0$, only for the negative detunings, $\delta_p=2\delta_s<0$, and that the high amplitude $q_-$ branch merges with the always unstable $q_+$ branch at some critical  detuning.  Examples of the numerically found soliton profiles and their associated combs for $P\ll P_{cr}$ are shown in Fig. 4.  The low pump solitons have  typical sech profiles. They are becoming narrower and the associated combs are becoming broader with the increased detuning as expected from Eq. (\ref{e4}).  Eq. (\ref{e4}) is not applicable for $P\sim P_{cr}$, since it completely disregards  Kerr nonlinearity. In this case, the soliton branches show a tendency to fold back together with the parametric cw's, see Figs. 1(b) and 1(d).   The two soliton branches do not join together as in the $P\ll P_{cr}$ case, but rather  terminate at a cw state, see the inset in Fig. 1(d). For the  relatively low intensities, these solitons resemble the ones for $P\ll P_{cr}$, see Fig. 5 (b).  However, when their intensities are high, the pulse profiles become the square-like connecting the low and high amplitude cw states via steep fronts, see Fig. 5(a). The associated  comb  develops the non-sech tails typical for the square pulses. In order to confirm that pulses emerging from the cw instabilities are indeed the soliton pulses, we have compared their amplitudes with  stationary soliton amplitudes for the both  $P\ll P_{cr}$ and $P\sim P_{cr}$ cases and found an impressive agreement, see  Fig. 6.


	Funding:	Leverhulme Trust (RPG-2015-456); Horizon 2020 Framework Programme (MICROCOMB: 812818); Russian Foundation for Basic Research (17-02-00081).

\end{document}